\documentclass[12pt]{article}
\usepackage{a4wide}

\usepackage[intlimits]{amsmath}
\usepackage{amstext,amsfonts,amsbsy,eucal,amssymb}
\usepackage{epsfig}
\usepackage{psfrag}

\usepackage{amsthm}
\usepackage{amscd}

\DeclareMathOperator{\tr}{tr}
\DeclareMathOperator{\str}{str}
\DeclareMathOperator{\fstr}{Str}
\begin{document}
\vspace*{30pt}
{\LARGE Fermionic ${R}$-Operator for the Fermion Chain Model} \\
\vspace{12pt}
\begin{center}
       Yukiko \textsc{Umeno}
        \footnote[2]{E-mail: 
                        \texttt{umeno@monet.phys.s.u-tokyo.ac.jp}
                   },
        Masahiro \textsc{Shiroishi}
        \footnote[8]{E-mail:
                        \texttt{siroisi@monet.phys.s.u-tokyo.ac.jp}
                   }   
        and
        Miki \textsc{Wadati}
\\
         Department of Physics, Graduate School of Science,\\
         University of Tokyo,\\
         Hongo 7-3-1, Bunkyo-ku, Tokyo 113-0033 
\end{center}

\begin{center}
(Received December 15, 1997)
\end{center}
\vspace{12pt}
\vspace{5mm}
The integrability of the one-dimensional (1D) fermion chain model is 
investigated in the framework of the Quantum Inverse Scattering Method
(QISM). We introduce a new $R$-operator for the fermion chain model,
which is expressed in terms of the fermion operators.
The $R$-operator satisfies a new type of the Yang-Baxter relation 
with fermionic $L$-operator. We derive the fermionic Sutherland
equation from the relation, which is equivalent to 
the fermionic Lax equation. It also provides a mathematical foundation
of the boost operator approach for the fermion model.
In fact, we obtain some higher conserved quantities 
of the fermion model using the boost operator. 
\vspace{3mm}
\begin{center}
{\bf KEYWORDS:}
\\
quantum inverse scattering method, \hspace{3mm}  fermionic formulation,\\
fermion chain model, \hspace{3mm} Sutherland equation, \hspace{3mm}
boost operator

\end{center}

\vspace{5mm}
\section{Introduction}

In the last decades, many integrable spin chain models have been
investigated by means of Quantum Inverse Scattering Method (QISM for brevity). 
\cite{Faddeev1}
In the QISM, the $R$-matrix, which satisfies the Yang-Baxter relation
with the $L$-operator
\begin{align}
R_{12}(u_1,u_2) \overset{1}{L}_j(u_1) \overset{2}{L}_j(u_2)
    = \overset{2}{L}_j(u_2) \overset{1}{L}_j(u_1) R_{12}(u_1,u_2),
\label{yang}
\end{align}
plays the essential role. \cite{Thacker,Kulish,Sklyanin1,Korepin,Faddeev2} 
For the spin chain models, 
the $R$-matrix $R_{12}(u_1,u_2)$ is a $c$-number matrix
and satisfies the Yang-Baxter equation
\begin{align}
R_{12}(u_1,u_2) R_{13}(u_1,u_3) R_{23}(u_2,u_3) 
     = R_{23}(u_2,u_3) R_{13}(u_1,u_3) R_{12}(u_1,u_2).
\end{align}
From the Yang-Baxter relation (\ref{yang}), it follows that 
the transfer matrix defined by  
\begin{align}
t(u) = \tr_a ( \overset{a}{L}_N(u) \ldots \overset{a}{L}_1(u) )
\end{align}
constitutes a commuting family 
\begin{align}
[ t(u) , t(v) ] = 0.
\end{align}
Expansion of $t(u)$ in terms of spectral parameter $u$
gives the conserved quantities. Therefore,
this proves the existence of the conserved operators $I^{(j)}$
which are involutive,
\begin{align}
[ I^{(j)} , I^{(k)} ] = 0.
\end{align}

We now know that there also exist many solvable fermion chain models
which are important in condensed matter physics. 
 They are usually
related to the spin models through the Jordan-Wigner transformation.
For example, the fermion chain model related to the $XXZ$ model  is
given by \cite{Pu,Wadati1}
\begin{align}
\mathcal{H} = \sum_{j=1}^N  \Bigl\{
             c_j^{\dagger} c_{j+1} + c_{j+1}^{\dagger} c_j  
           + \dfrac{\Delta}{2} (2 n_j - 1)(2 n_{j+1} - 1) 
           \Bigr\}.
\label{fxxz}
\end{align}
Here $c_j^{\dagger}$ and $c_j$ are the creation and the annihilation
operators at site $j$ which satisfy the canonical anti-commutation relations
\begin{align}
\{ c_j ,c_k \} = \{ c_j^{\dagger} , c_k^{\dagger} \} = 0, 
   \hspace{5mm}  \{ c_j , c_k^{\dagger} \} = \delta_{jk},
\end{align}
and $n_j$ is the number operator
\begin{align}
n_j = c_j^{\dagger} c_j.
\end{align}
The periodic boundary condition (PBC) is assumed in (\ref{fxxz}),
\begin{align}
c_{N+1} = c_1, \hspace{5mm} c_{N+1}^{\dagger} = c_1^{\dagger}.
\label{pbc}
\end{align}
We refer to this model (\ref{fxxz}) as the ${XXZ}$ fermion model.

Since the Jordan-Wigner transformation connects the spin model and the 
fermion model, physical properties of the fermionic chain models 
including the exact integrability are often discussed using 
the corresponding spin chain model.
However, there are some important  differences between the fermion models and
the spin models. In particular, the PBC for the fermion model does not
correspond to the PBC for the spin model due to the non-locality of the 
Jordan-Wigner transformation. Therefore it is more appropriate to treat
the fermion models keeping their fermionic nature as much as possible
(c.f. refs. 9 and 10).

The QISM for the fermion chain model (\ref{fxxz}) has been developed by
several authors, \cite{Pu,Wadati1,Wadati2,Zhou}
where the fermionic ${L}$-operator is introduced.
The fermionic $R$-matrix which 
satisfies the graded Yang-Baxter relation was also found. 
The existence of the fermionic $R$-matrix can 
be used to establish the integrability of the fermion model under the
PBC (\ref{pbc}) or other twisted boundary conditions. 
The most intriguing point of this
method is that the matrix elements of the fermionic $L$-operator
consist of the fermion operators, which means that the quantum space
is fermionic. On the other hand, the auxiliary space remains to be a
usual spin space and correspondingly the $R$-matrix has the $c$-number 
elements. The fact suggests that there still requires further 
investigation on the fermionic formulation of the QISM.

In this paper, we introduce an $R$-operator for the $XXZ$ fermion
model. It consists of the fermion operators and satisfies a new type
of the Yang-Baxter relation with the fermionic $L$-operator.
The role of the auxiliary space and the quantum space for the
fermionic $L$-operator is exchanged in the Yang-Baxter relation. We
derive the fermionic Sutherland equation from the Yang-Baxter relation, 
which is shown to be equivalent to the fermionic Lax
equation. \cite{Zhou,Wadati0}  To obtain the higher conserved 
operators for the lattice spin models, it is in general very useful to 
introduce the boost operator. \cite{Tetelman,Sogo,Mathieu} 
The fermionic Sutherland equation 
provides the basis for the application
of the boost operator to the fermion models.  
We derive some higher
conserved operators for the ${XXZ}$ fermion model using the boost operator.

The paper is organized as follows. In \S 2, we summarize salient
points of the QISM
for the $XXZ$ fermion model. In \S 3, we consider the exchange of 
the auxiliary and quantum spaces of the fermionic $L$-operator and 
introduce an $R$-operator. We also discuss
the fundamental properties of the $R$-operator. In particular, we find 
that the $R$-operator satisfies the Yang-Baxter relation. 
In \S 4, we further discuss some applications of the fermionic
$R$-operator and the Yang-Baxter relation. We derive the Sutherland
equation and introduce the boost operator.
The last section is devoted to concluding remarks.

\section{Graded Yang-Baxter Relation for the $XXZ$ Fermion Model}

In this section we briefly summarize the QISM for the $XXZ$ fermion model
(\ref{fxxz}). The fermionic $L$-operator is given by  
\begin{align}
\mathcal{L}_j(u) =
    \begin{pmatrix}
      \mbox{i} \alpha(u) n_j + \gamma(u) (1-n_j) & 2\beta(u) c_j\\
      -2\mbox{i} \beta(u) c_j^{\dagger} & 
          \alpha(u) (1-n_j) - \mbox{i} \gamma(u) n_j
    \end{pmatrix},
\label{ferl}
\end{align}
where
\begin{align}
\alpha(u)&=\sin(u+2\eta),
\nonumber\\
\gamma(u)&=\sin u,
\nonumber\\
2\beta(u)&=\sin 2\eta. 
\end{align}
We express by $\underset{s}{\otimes}$ the  Grassmann (graded) direct
product
\begin{align} 
[ A \underset{s}{\otimes} B ]_{\alpha \gamma , \beta \delta} 
  &= (-1)^{[P(\alpha) + P(\beta)] P(\gamma)} 
      A_{\alpha \beta} B_{\gamma \delta}
\nonumber\\
P(1) &= 0, \hspace{5mm} P(2) = 1.
\end{align}
Then there exists the fermionic $R$-matrix which satisfies the graded
Yang-Baxter relation (Fig.\ 1),
\begin{align}
\mathcal{R}_{12}(u_1-u_2) 
 \overset{1}{\mathcal{L}}(u_1) \overset{2}{\mathcal{L}}(u_2)
  = \overset{2}{\mathcal{L}}(u_2) \overset{1}{\mathcal{L}}(u_1)
     \mathcal{R}_{12}(u_1-u_2) 
\label{gyangl}
\end{align}
where, with $\mathcal{I}$ being 2 $\times$ 2 identity matrix,
\begin{align}
\overset{1}{\mathcal{L}}(u_1) 
= \mathcal{L}_j(u_1) \underset{s}{\otimes} \mathcal{I},
\hspace{5mm}
\overset{2}{\mathcal{L}}(u_2) 
= \mathcal{I} \underset{s}{\otimes} \mathcal{L}_j(u_2).
\end{align}
The explicit form of the fermionic $R$-matrix is 
\begin{align} 
\mathcal{R}_{12}(u) =
    \begin{pmatrix} 
      \sin (u+2\eta) & 0 & 0 & 0\\
      0 & - \mbox{i}\sin u & \sin 2\eta & 0\\
      0 & \sin 2\eta & \mbox{i}\sin u & 0\\
      0 & 0 & 0 & - \sin (u+2\eta)
    \end{pmatrix}.
\label{gR}
\end{align}
It fulfills the graded Yang-Baxter equation (Fig.\ 2)
\begin{align} 
\mathcal{R}_{12}(u) \mathcal{R}_{13}(u+v) \mathcal{R}_{23}(v) 
 = \mathcal{R}_{23}(v) \mathcal{R}_{13}(u+v) \mathcal{R}_{12}(u) 
\label{gyang}
\end{align}
Note that non-zero elements of (\ref{gR}) are even with respect
to the parity $P(\alpha)$, i.e.,
\begin{align} 
P(\alpha) + P(\beta) + P(\alpha') + P(\beta') = 0  
                   \hspace{5mm} (\mbox{mod 2})
                 \hspace{5mm}  \mbox{for} \hspace{5mm}
               \mathcal{R}_{\alpha \beta , \alpha' \beta'} \neq 0 .
\end{align}

The monodromy matrix $T(u)$ is defined as an ordered product of the fermionic
$L$-operators
\begin{align} 
T(u) = \mathcal{L}_N(u) \ldots \mathcal{L}_1(u).
\end{align}
From the local relation (\ref{gyangl}), we have the global relation
for the monodromy matrix
\begin{align} 
\mathcal{R}_{12}(u_1-u_2) 
\overset{1}{T}_j(u_1) \overset{2}{T}_j(u_2)
 = \overset{2}{T}_j(u_2) \overset{1}{T}_j(u_1) 
\mathcal{R}_{12}(u_1-u_2) . 
\label{gyangt}
\end{align}

The transfer matrix $\tau (u)$ is defined by the supertrace of the
monodromy matrix $T(u)$ 
\begin{align} 
\tau (u) = \str T(u) \equiv \tr \{ \sigma^z T(u) \} .
\end{align}
Then the global graded Yang-Baxter relation (\ref{gyangt}) leads to
the commutativity of the transfer matrix, 
\begin{align} 
[ \tau (u_1) , \tau (u_2) ] = 0 .
\end{align}
The last equation establishes the exact integrability of the model
(\ref{fxxz}) under the PBC. In fact, we have 
\begin{align} 
\tau (u) = \tau(0) \{ 1 + u \mathcal{H} + \ldots \} ,
\end{align}
with the identification
\begin{align} 
\Delta = \cos 2\eta .
\end{align}
For later use, we introduce a notation
\begin{align} 
\mathcal{H} = \sum_{j=1}^N \mathcal{H}_{j,j+1} ,
\end{align}
where
\begin{align} 
\mathcal{H}_{j,k} = \dfrac{1}{\sin 2\eta} 
            \Bigl\{ ( c^{\dagger}_j c_k + c^{\dagger}_k c_j )
                + \dfrac{1}{2} \cos 2\eta (2n_j-1)(2n_k-1)
                + \frac{1}{2} \cos 2\eta
             \Bigr\} .
\label{2pham}
\end{align}
We call $\mathcal{H}_{j,k}$ the two-point Hamiltonian density.

\vspace{8mm}
\begin{small}
  \begin{equation*}
    \begin{psfrags}
      \psfrag{EQ}{$=$} 
      \psfrag{1}{1}
      \psfrag{2}{2}
      \psfrag{3}{j}
      \psfig{file=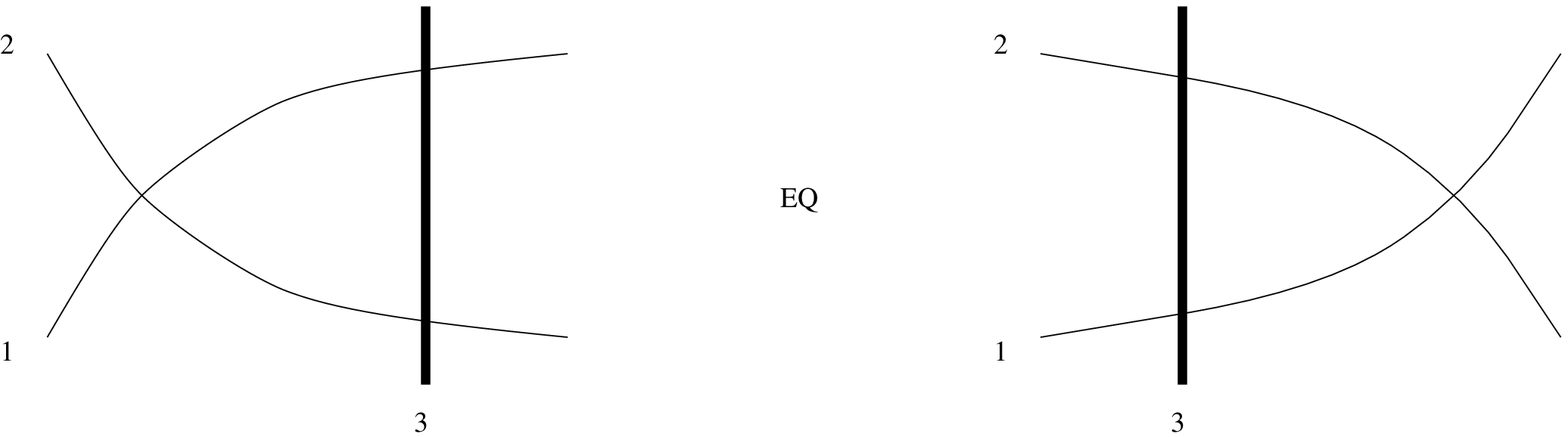,scale=0.6}
\end{psfrags}  
\end{equation*}   
\end{small}
\begin{center}
{\bf Fig.\ 1  Graded Yang-Baxter relation, eq.\ (13)}
\end{center}
  
 \begin{small}
 \begin{equation*}
    \begin{psfrags}
      \psfrag{EQ}{$=$}
      \psfrag{1}{1}
      \psfrag{2}{2}
      \psfrag{3}{3}  
      \psfig{file=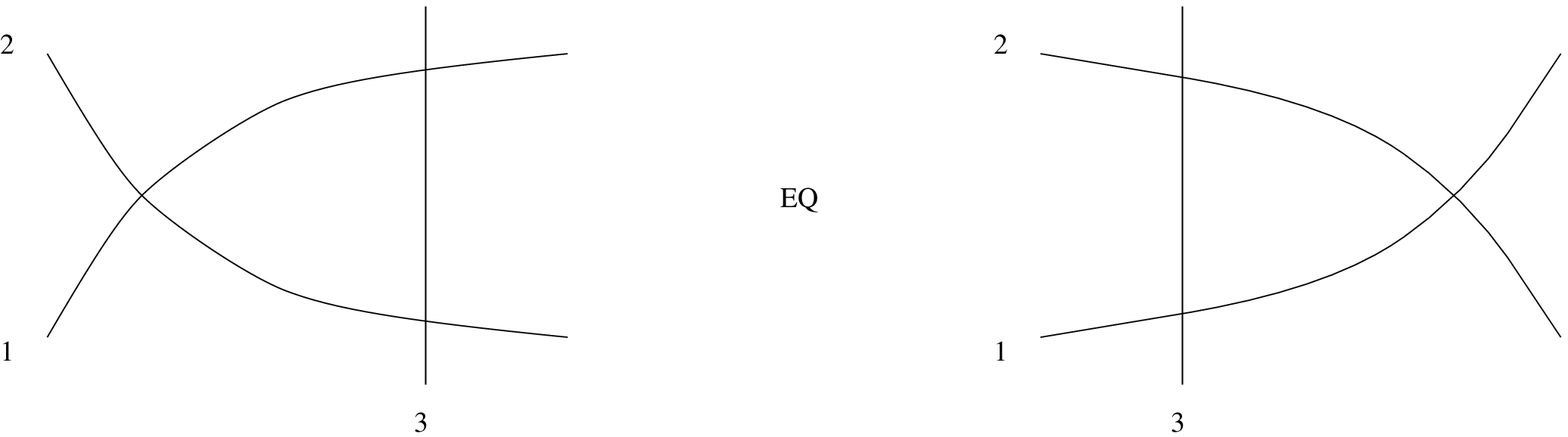,scale=0.6}  
\end{psfrags}  
\end{equation*}
\end{small}

\begin{center}
{\bf Fig.\ 2  Graded Yang-Baxter equation, eq.\ (16)}
\end{center}

\section{Fermionic $R$-Operator}

One of the characteristic features of the fermionic $L$-operator (\ref{ferl})
is that its matrix elements consist of the fermion operators, that is, 
quantum space of the fermionic $L$-operator is fermion Fock 
space. However, the auxiliary space remains to be a usual spin space.
The fermionic $L$-operator is therefore asymmetric with respect to
the exchange of the quantum space and the auxiliary space. The reader
may recall the similar situation when we construct the spin-$S$ $XXZ$
model from the spin-$\frac{1}{2}$ $XXZ$ model. In this case, we first
consider the intermediate  $L$-operator, whose quantum space 
consists of spin operators of magnitude $S$. Then we exchange the
auxiliary and quantum spaces of the $L$-operator and consider a new
Yang-Baxter relation. The $R$-matrix which satisfies the
Yang-Baxter relation should be in a $(2S+1) \times (2S+1)$ matrix
form.  From the 
$R$-matrix, we can construct a new transfer matrix which generates the 
spin-$S$ $XXZ$ model. \cite{Faddeev2,Sogo1,Gomez}

We apply the similar procedure to the fermion chain model.
Consider the following new Yang-Baxter relation (Fig.\ 3)
\begin{align} 
\mathcal{R}^{\rm f}_{12}(u_1-u_2) \overset{a}{\mathcal{L}}_1(u_1) 
     \overset{a}{\mathcal{L}}_2(u_2)
   =\overset{a}{\mathcal{L}}_2(u_2) \overset{a}{\mathcal{L}}_1(u_1) 
       \mathcal{R}^{\rm f}_{12}(u_1-u_2)
 \label{fyangl}
\end{align}

Notice that the role of the auxiliary and quantum spaces of the fermionic
$L$-operator is exchanged in (\ref{gyangl}) and (\ref{fyangl}). 
Inevitably, the $R$-operator
$\mathcal{R}^{\rm f}(u)$ is no longer $c$-number matrix and consists of
fermion operators. Of course, there is no assurance for the existence
of such $R$-operator at this stage. We assume a form of the
$R$-operator as
\begin{align} 
\mathcal{R}^{\rm f}_{12}(u_1-u_2)=&
            g_1 n_1 n_2 + g_2 n_1 (1-n_2)
\nonumber\\
            & + g_3 (1-n_1) n_2 + g_4 (1-n_1) (1-n_2)
\nonumber\\
            & + g_5 c_1^{\dagger} c_2 + g_6 c_2^{\dagger} c_1 ,         
\label{ferr}
\end{align}
where $g_i$, $i = 1, \ldots ,6$ are scalar functions of $u_1$ and
$u_2$. Substituting the expression (\ref{ferr}) in (\ref{fyangl}),
we have the following equations,
\begin{align} 
        g_4 &= - g_1, \hspace{3mm} g_3 = g_2, \hspace{3mm} g_6 = g_5,
\nonumber\\
        \alpha (u_2) g_1 &= - \gamma (u_2) g_2 - \alpha (u_1) g_5,
\nonumber\\
        \gamma (u_1) g_1 &= - \alpha (u_1) g_2 - \gamma (u_2) g_5,
\nonumber\\
        \beta (u_1) \beta (u_2) g_2 &= \{  \gamma (u_1) \alpha (u_2)
                                  - \alpha (u_1) \gamma (u_2)  \} g_5. 
\label{relg}
\end{align}
A set of functional equations (\ref{relg}) is solved to give 
(up to an overall factor) 
\begin{align} 
       g_1 &= - g_4 = - \dfrac{\sin (u_1-u_2+2\eta)}{\sin 2\eta},
\nonumber\\
       g_2 &= g_3 = \dfrac{\sin (u_1-u_2)}{\sin 2\eta},
\nonumber\\
       g_5 &= g_6 = 1.
\end{align}
Thus obtained $R$-operator, which we call fermionic $R$-operator,
\begin{align} 
       \mathcal{R}^{\rm f}_{12} (u_1-u_2) =& 
               \dfrac{\sin (u_1-u_2+2\eta)}{\sin 2\eta} (1 - n_1 - n_2)
\nonumber\\
               &  + \dfrac{\sin (u_1-u_2)}{\sin 2\eta} (n_1 + n_2 - 2 n_1 n_2)
\nonumber\\
               & + c_1^{\dagger} c_2 + c_2^{\dagger} c_1 ,         
\end{align}
satisfies (\ref{fyangl}).

The fermionic $R$-operator enjoys the following properties.
\\
1. Regularity (Initial condition)
\begin{align} 
        \mathcal{R}^{\rm f}_{12}(u=0) = \mathcal{K}_{12}.
\end{align}
Here
\begin{align}
        \mathcal{K}_{ij} = 1 - (c_i^{\dagger}-c_j^{\dagger})(c_i-c_j)
\end{align}
is the permutation operator for the fermion operators:
\begin{align} 
        \mathcal{K}_{ij} &= \mathcal{K}_{ji},  \hspace{5mm}
        \mathcal{K}_{jj} = 1,
\nonumber\\
        \mathcal{K}_{ij} c_i &= c_j \mathcal{K}_{ij} , \hspace{5mm}
        \mathcal{K}_{ij} c_i^{\dagger} = c_j^{\dagger} \mathcal{K}_{ij} ,
\nonumber\\
        \mathcal{K}_{ij} \mathcal{K}_{ij} &= 1 .       
\end{align}
2. Local Hamiltonian:
\begin{align} 
       \dfrac{{\rm d}\mathcal{R}^{\rm f}_{12}(u)}{{\rm d} u} \Big|_{u=0} 
                    = \mathcal{K}_{12} \mathcal{H}_{12}.    
\end{align}       
Here $\mathcal{H}_{12}$ is the 2-point Hamiltonian density given in
(\ref{2pham}).
\\
3. Yang-Baxter equation:

The fermionic $R$-operator satisfies the Yang-Baxter equation (Fig.\ 4) as 
\begin{align} 
\mathcal{R}^{\rm f}_{12}(u_1-u_2) \mathcal{R}^{\rm f}_{13}(u_1) \mathcal{R}^{\rm f}_{23}(u_2) = 
   \mathcal{R}^{\rm f}_{23}(u_2) \mathcal{R}^{\rm f}_{13}(u_1) \mathcal{R}^{\rm f}_{12}(u_1-u_2).
\label{fyang}
\end{align}
4. Unitarity:
\begin{align}
        \mathcal{R}^{\rm f}_{12}(u) \mathcal{R}^{\rm f}_{12}(-u) 
             = \Bigl[ 1- \dfrac{\sin ^2 u}{\sin ^2 2\eta} \Bigr] \mathbf{1}.
\end{align}

Now we consider a product of the fermionic $R$-operators
\begin{align} 
\mathcal{T}^{\rm f}_a(u) 
=  \mathcal{R}^{\rm f}_{aN}(u) \ldots \mathcal{R}^{\rm f}_{a2}(u) 
                         \mathcal{R}^{\rm f}_{a1}(u).
\end{align}
Then from (\ref{fyang}), we see that $\mathcal{T}^{\rm f}_a(u)$ satisfies the
Yang-Baxter relation,  
\begin{align} 
\mathcal{R}^{\rm f}_{12}(u_1-u_2) \mathcal{T}^{\rm f}_1(u_1) \mathcal{T}^{\rm f}_2(u_2)
       = \mathcal{T}^{\rm f}_2(u_2) \mathcal{T}^{\rm f}_1(u_1 )\mathcal{R}^{\rm f}_{12}(u_1-u_2).
\label{fyangt}
\end{align}
Here we note the commutativity,
\begin{align} 
\mathcal{R}^{\rm f}_{1j}(u_1) \mathcal{R}^{\rm f}_{2k}(u_2)
             = \mathcal{R}^{\rm f}_{2k}(u_2) \mathcal{R}^{\rm f}_{1j}(u_1), 
              \hspace{5mm} (j \neq k).
\end{align}
We refer to $\mathcal{T}^{\rm f}_j(u)$ as monodromy operator. We can also
introduce an analog of the transfer matrix,
\begin{align} 
        \tau ^{\rm f}(u) &= \fstr \mathcal{T}^{\rm f}_a(u)
\nonumber\\
       & \equiv   {}_a \langle 0 | \mathcal{T}^{\rm f}_a(u) | 0 \rangle _a 
                 - {}_a \langle 1 | \mathcal{T}^{\rm f}_a(u) | 1 \rangle _a , 
\end{align}
where $ | 0 \rangle _a$ and $ | 1 \rangle _a$ are defined by 
\begin{align} 
           c_a | 0 \rangle _a = 0 ,  \hspace{5mm}            
           | 1 \rangle _a = c_a^{\dagger} | 0 \rangle _a . 
\end{align}
The fermionic transfer operator $\tau^{\rm f}(u)$ constitutes 
a commuting family due to 
(\ref{fyangt})
\begin{align} 
    [ \tau ^{\rm f}(u) , \tau ^{\rm f}(v) ] = 0,
\end{align}
which also gives Hamiltonian (\ref{fxxz}),
\begin{align}
    \mathcal{H} = \sum_{j=1}^N \mathcal{H}_{j,j+1}
                = \dfrac{{\rm d} \log \tau^{\rm f}(u)}{{\rm d} u} \Big|_{u=0}.
\end{align}
\vspace{8mm}
  \begin{small}
  \begin{equation*}
    \begin{psfrags}
      \psfrag{EQ}{$=$} 
      \psfrag{1}{1}
      \psfrag{2}{2}
      \psfrag{3}{a}
     \psfig{file=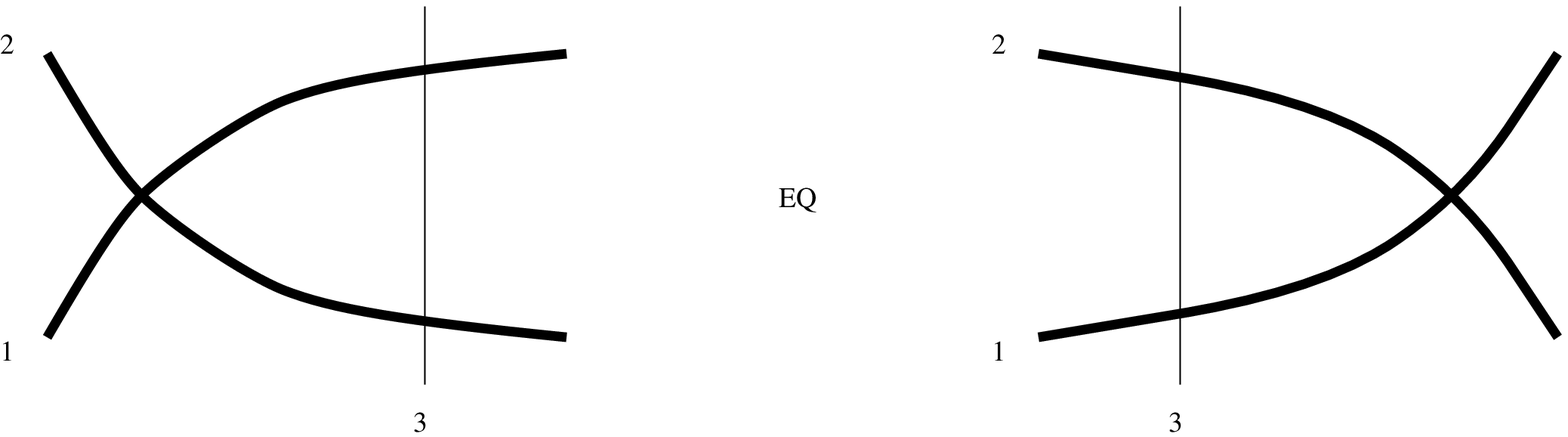,scale=0.6}
    \end{psfrags}  
\end{equation*}
\end{small}
\begin{center}
{\bf Fig.\ 3  Fermionic Yang-Baxter relation, eq.\ (26)}
\end{center}

  \begin{small}
  \begin{equation*}
    \begin{psfrags}
      \psfrag{EQ}{$=$} 
      \psfrag{1}{1}
      \psfrag{2}{2}
      \psfrag{3}{3}
      \psfig{file=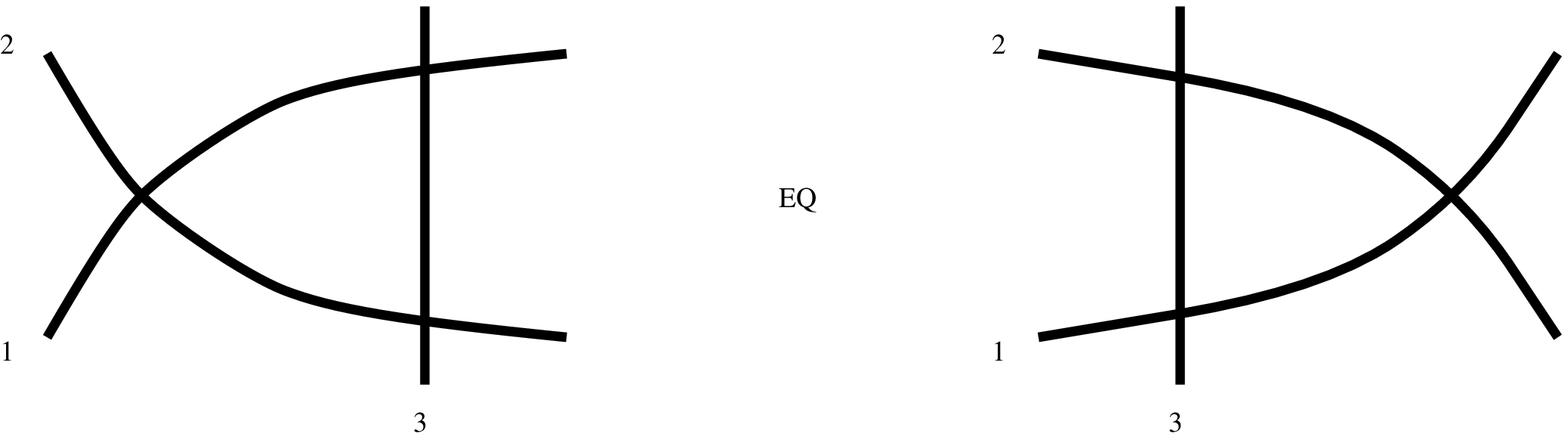,scale=0.6}
    \end{psfrags}  
\end{equation*}
\end{small}
\begin{center}
{\bf Fig.\ 4  Fermionic Yang-Baxter equation, eq.\ (35)}
\end{center}

\section{Sutherland Equation and Boost Operator}

In this section we discuss further some applications of the fermionic
$R$-operator and the Yang-Baxter relation.

First, we differentiate the Yang-Baxter relation (\ref{fyangl}) 
with respect to the
spectral parameter $u_2$ at $u_2 = u_1$. Then we get
\begin{align} 
 - \mathcal{K}_{12} \mathcal{H}_{12} 
            \mathcal{L}_1(u_1) \mathcal{L}_2(u_1) 
  + \mathcal{K}_{12} \mathcal{L}_1(u_1) 
     \dfrac{{\rm d} \mathcal{L}_2(u_2)}{{\rm d} u_2} \Bigl| _{u_2=u_1}
\nonumber\\
 = \dfrac{{\rm d} \mathcal{L}_2(u_2)}{{\rm d} u_2} \Bigl| _{u_2=u_1}
             \mathcal{L}_1(u_1) \mathcal{K}_{12}           
  - \mathcal{L}_2(u_1) \mathcal{L}_1(u_1) 
            \mathcal{K}_{12} \mathcal{H}_{12}. 
\label{dfyangl}
\end{align}
For simplicity, we do not write the auxiliary space dependence of the
$L$-operator here and hereafter. By use of the relations 
\begin{align}
       \mathcal{L}_1(u) \mathcal{K}_{12}
             =  \mathcal{K}_{12} \mathcal{L}_2(u),
        \hspace{5mm}
       \mathcal{L}_2(u) \mathcal{K}_{12}        
             =  \mathcal{K}_{12} \mathcal{L}_1(u), 
\end{align} 
eq.\ (\ref{dfyangl}) can be rewritten as
\begin{align}
 [ \mathcal{H}_{12} , \mathcal{L}_1(u) \mathcal{L}_2(u) ]  
   = \mathcal{L}_1(u) \dfrac{{\rm d} \mathcal{L}_2}{{\rm d} u} (u)   
       - \dfrac{{\rm d} \mathcal{L}_1}{{\rm d} u} (u) \mathcal{L}_2(u),
\label{fsuth}
\end{align} 
This is the fermionic version of the Sutherland equation, which
enables us   
to show the commutativity of the Hamiltonian
$\mathcal{H}$ and the transfer matrix $\tau (u)$,
\begin{align} 
 [ \mathcal{H} , \tau (u) ] = 0.
\end{align}
The fermionic Sutherland equation (\ref{fsuth}) is equivalent to the
Lax equation,
\begin{align}
     \dfrac{{\rm d} \mathcal{L}_j(u)}{{\rm d} t}
       & \equiv \mbox{i} [ \mathcal{L}_j(u) , \mathcal{H} ]
\nonumber\\
       & = \mathcal{M}_{j+1}(u) \mathcal{L}_j(u) 
                   - \mathcal{L}_j(u) \mathcal{M}_j(u).
\label{flax}
\end{align} 
Here $\mathcal{M}_j$ is given by the formula, 
\begin{align} 
            \mathcal{M}_j(u) = - \mbox{i} \mathcal{L}_j^{-1}(u)
                \Bigl\{  \dfrac{{\rm d} \mathcal{L}_j(u)}{{\rm d} u}   
                 + [ \mathcal{H}_{j,j-1} , \mathcal{L}_j(u) ]  \Bigr\}.
\label{m}
\end{align}
We remark that the formula (\ref{m}) for the usual spin models is
well-known. \cite{Korepin} However, to our knowledge, the formula (\ref{m}) for the fermion models is not known before and is proved here for the first time. We can conclude that the fermionic Lax equation
follows from the Yang-Baxter relation (\ref{fyangl}).

Using (\ref{ferl}) and (\ref{2pham}) we can obtain 
the Lax operator $\mathcal{M}_j(u)$ for the
fermion $XXZ$ model as 
\begin{align}
       \mathcal{M}_j(u) = 
    \begin{pmatrix}
        \mathcal{M}_{11}(u) & \mathcal{M}_{12}(u) \\
        \mathcal{M}_{21}(u) & \mathcal{M}_{22}(u)
    \end{pmatrix} ,
\end{align}
where
\begin{align}
       \mathcal{M}_{11}(u) =& 
             \dfrac{1}{\sin 2\eta} 
                    \Bigl\{ \mbox{i} - \dfrac{\sin u}{\sin (u+2\eta)} \Bigr\}
                         c_j^{\dagger} c_{j-1}
           + \dfrac{1}{\sin 2\eta}
                    \Bigl\{ \mbox{i} + \dfrac{\sin u}{\sin (u-2\eta)} \Bigr\}
                         c_{j-1}^{\dagger} c_j
\nonumber\\
           &- \dfrac{\mbox{i} \sin 4\eta}{\sin (u+2\eta) \sin (u-2\eta)}
                         (1-n_j)(1-n_{j-1})
           - \mbox{i} \cot (u+2\eta) ,
\nonumber\\
       \mathcal{M}_{12}(u) =& 
                  \dfrac{1}{\sin (u-2\eta)}
                              \{ \mbox{i} n_{j-1} c_j - (1-n_j) c_{j-1} \}   
                -  \dfrac{1}{\sin (u+2\eta)}
                              \{ \mbox{i} (1-n_{j-1}) c_j - n_j c_{j-1} \} ,  
\nonumber\\
       \mathcal{M}_{21}(u) =& 
                - \dfrac{1}{\sin (u+2\eta)}
          \{ \mbox{i} n_{j-1} c_j^{\dagger} + (1-n_j) c_{j-1}^{\dagger} \}   
                + \dfrac{1}{\sin (u-2\eta)}
          \{ \mbox{i} (1-n_{j-1}) c_j^{\dagger} + n_j c_{j-1}^{\dagger} \} ,   
\nonumber\\
       \mathcal{M}_{22}(u) =& 
             \dfrac{1}{\sin 2\eta} 
                    \Bigl\{ \mbox{i} - \dfrac{\sin u}{\sin (u+2\eta)} \Bigr\}
                         c_{j-1}^{\dagger} c_j
           + \dfrac{1}{\sin 2\eta} 
                    \Bigl\{ \mbox{i} + \dfrac{\sin u}{\sin (u+2\eta)} \Bigr\}
                         c_j^{\dagger} c_{j-1}
\nonumber\\
           &- \dfrac{\mbox{i} \sin 4\eta}{\sin (u+2\eta) \sin (u-2\eta)}
                         n_j n_{j-1}
           - \mbox{i} \cot (u+2\eta) .
\end{align} 
This result coincides with the known one \cite{Zhou} up to the terms
proportional to the identity matrix. 

As another application of the fermionic Sutherland equation, we 
consider the boost operator for the fermion chain model as 
\begin{align}
     \mathcal{B} = \sum _{j=1}^N j \mathcal{H}_{j,j+1}.
\label{boost}
\end{align} 
By making use of the fermionic Sutherland equation, we can show the
following relations,
\begin{align} 
   [ \mathcal{B} , \tau (u) ] = \dfrac{{\rm d} \tau (u)}{{\rm d} u} 
              - N \str   \Bigl(  \mathcal{L}_N \ldots \mathcal{L}_2 
                      \dfrac{{\rm d} \mathcal{L}_1}{{\rm d} u}
                                     \Bigr).
\label{btau}
\end{align}
We define  $I^{(n)}$ as $n$-th local conserved operator and $I^{(1)} =
\mathcal{H}$. Then, from (\ref{btau}), a recursion relation 
for the conserved operator is
\begin{align}
        I^{(n+1)} = [ \mathcal{B} , I^{(n)}] 
                           + (\mbox{some other terms}).
\label{rec}
\end{align}
By use of the boost operator (\ref{boost}) and 
the recursion relation (\ref{rec}), we immediately obtain some 
higher conserved operators for the fermion $XXZ$ model.
Notice that the second term in the r.h.s. of eq.\ (\ref{rec}) 
originates from the second term in the r.h.s. of eq.\ (\ref{btau}),   
and 
we can estimate its contribution.
Then we get
\begin{align}
   I^{(2)} =& \dfrac{1}{\sin ^2 2\eta} \sum_{j=1}^N  \Bigl[
          (c_{j+2}^{\dagger} c_j - c_j^{\dagger} c_{j+2})
          -2 \cos 2\eta   \Bigl\{
           (c_{j+1}^{\dagger} c_j - c_j^{\dagger} c_{j+1})
\nonumber\\
          &- n_{j+2} (c_{j+1}^{\dagger} c_j - c_j^{\dagger} c_{j+1})
          - (c_{j+2}^{\dagger} c_{j+1} - c_{j+1}^{\dagger} c_{j+2}) n_j
                    \Bigr \}   \Bigr]
\end{align} 
\begin{align}
    I^{(3)} =& \dfrac{2}{\sin ^3 2\eta}   
                   \sum_{j=1}^N   \Bigl[
       ( c_{j+3}^{\dagger} c_j + c_j^{\dagger} c_{j+3}
            + c_{j+1}^{\dagger} c_j + c_j^{\dagger} c_{j+1} )
\nonumber\\
      &+ \cos 2\eta       \Bigl\{
            -3 (c_{j+2}^{\dagger} c_j + c_j^{\dagger} c_{j+2})
            + 2 n_{j+3} (c_{j+2}^{\dagger} c_j + c_j^{\dagger}c_{j+2}) 
\nonumber\\
            &+ 2 n_{j+1} (c_{j+2}^{\dagger} c_j + c_j^{\dagger}c_{j+2})         
            + 2 n_{j-1} (c_{j+2}^{\dagger} c_j + c_j^{\dagger}c_{j+2})         
\nonumber\\
            &+ 2 (c_{j+3}^{\dagger} c_{j+2} - c_{j+2}^{\dagger}c_{j+3})
                   (c_{j+1}^{\dagger} c_j - c_j^{\dagger} c_{j+1})
            - 2 n_j + 4 n_{j+1} n_j - 2 n_{j+2} n_j      \Bigr\}
\nonumber\\        
       &+ 2 \cos ^2 2\eta      \Bigl\{
              (c_{j+1}^{\dagger} c_j + c_j^{\dagger} c_{j+1})
            + 2 n_{j+3}
               (c_{j+2}^{\dagger} c_{j+1} + c_{j+1}^{\dagger} c_{j+2}) n_j 
\nonumber\\
            &- n_{j+2} (c_{j+1}^{\dagger} c_j + c_j^{\dagger} c_{j+1})
            - n_j (c_{j+2}^{\dagger} c_{j+1} + c_{j+1}^{\dagger} c_{j+2})
                            \Bigr\}    \Bigr] .
\end{align} 
We remark that these results are consistent with the known ones. 
\cite{Wadati2,Zhou,Zhou1}

\section{Concluding Remarks}

In this paper, we have studied the quantum inverse scattering method (QISM)
for the $XXZ$ fermion model.
We have introduced a fermionic $R$-operator for the fermion
chain model. It consists of the fermion operators and satisfies the
Yang-Baxter equation. The  $R$-operator intertwines the fermionic
$L$-operator in a different way from the usual one. The auxiliary
space and the quantum space of the fermionic $L$-operator are
exchanged. We have shown that 
the new relation plays a complementary role to the usual
graded Yang-Baxter relation. In particular, we can derive the fermionic
Sutherland equation from the relation. It is shown that the fermionic
Lax equation is equivalent to the fermionic Sutherland equation. 
The Sutherland equation 
also gives a mathematical foundation of the boost operator approach 
for the fermion chain model.

We have also defined a fermionic monodromy operator by a product of 
the fermionic  $R$-operators. The transfer operator, which is an analog 
of the transfer matrix, can be constructed by taking the expectation 
values in the auxiliary space, $\fstr$, of the
monodromy operator. The transfer operator constitutes a commuting
family, which proves the integrability of the fermion chain model. The
diagonalization of the fermionic transfer operator will be discussed
in a separate paper.

Our approach is applicable to other integrable fermion models. 
Among them, we
shall consider the Hubbard model in subsequent papers.

\begin{center}
{\bf Acknowledgment}
\end{center}

The authors are grateful to K. Hikami, H. Ujino, Y. Komori, 
T. Tsuchida and R. Inoue for useful discussions and careful 
reading of the manuscript. This work is in part supported by 
Grant-in-Aid for JSPJ Fellows from the Ministry of Education, 
Science, Sports and Culture of Japan.


\end{document}